\documentclass[conference]{IEEEtran}

\usepackage{graphicx,times,amsmath}
\usepackage[lofdepth,lotdepth]{subfig}

\IEEEoverridecommandlockouts

\begin{document}

\title{\ \\ \LARGE\bf Application of a clustering framework to UK domestic electricity data 
\thanks{Corresponding author is Ian Dent (phone: +44 115 846 6568; email: {\tt
ird@cs.nott.ac.uk}), University of Nottingham, School of Computer Science, Jubilee Campus, Nottingham, NG8 1BB, UK}
\thanks{Professor Uwe Aickelin is with the Intelligent Modelling and Analysis Group, University of Nottingham, (email: {\tt
uwe.aickelin@nottingham.ac.uk}).}
\thanks{Professor Tom Rodden is with the Horizon Digital Economy Research Institute, University of Nottingham, (email: {\tt
tar@cs.nott.ac.uk}).}}

\author{Ian Dent, Uwe Aickelin, Tom Rodden}

\maketitle

\begin{abstract}
This paper takes an approach to clustering domestic electricity load profiles that has been successfully used with data from Portugal and applies it to UK data. Clustering techniques are applied and it is found that the preferred technique in the Portuguese work (a two stage process combining Self Organised Maps and Kmeans) is not appropriate for the UK data. The work shows that up to nine clusters of households can be identified with the differences in usage profiles being visually striking. This demonstrates the appropriateness of breaking the electricity usage patterns down to more detail than the two load profiles currently published by the electricity industry.

The paper details initial results using data collected in Milton Keynes around 1990. Further work is described and will concentrate on building accurate and meaningful clusters of similar electricity users in order to better direct demand side management initiatives to the most relevant target customers.

\end{abstract}


\section{Introduction}

The electricity market in the UK is currently undergoing a period of major change and is being subjected to various pressures. Some of these pressures are arising from UK specific situations, such as the history and current design of the National Grid, and others from worldwide trends, such as the need to reduce carbon emissions and the declining sources of hydro-carbon fuels. New technologies, such as electric cars which will need household charging facilities, are expected to become much more prevalent. The information available to monitor and to manipulate the electricity usage will grow very rapidly, particularly with the roll out of Smart Meters which is planned to be complete in the UK by 2019. In addition, the drive to change the mix of electricity generation technologies in order to reduce greenhouse gas emissions, the desire to reduce carbon dioxide by switching non-electric demand such as gas central heating to the electricity network, and the impact of climate change, with its associated change in electricity demand and more frequent extreme weather events, will impact on the market.

There is a trend towards providing more targeted and more complicated tariff offers for customers in order to provide many benefits including maximising the efficiency of the supply process. \cite{energy2009} shows that the provision of Smart Meters will allow greatly increased analysis of a customer's electricity usage and provide the ability to make customised offers on pricing and supply availability. This will offer an opportunity to change customer behaviour (for example, to minimise usage during peak periods) or to increase efficiencies in the electricity supply chain in meeting the predicted demand 
\cite{ofgem2010}.

The identification of typical electrical usage patterns within households is necessary as a starting point for:
\begin{itemize}
\item{Defining the type of Demand Side Management program (e.g. peak clipping) to undertake to match the overall electricity supply goals.}
\item{Assessing the impact of any initiatives to reduce overall energy usage in order to discover the amount of overall reduction which occurs during different times of the day.}
\item{Allowing accurate aggregation to provide a pattern of total demand to be met by supply side generation and transmission.}
\end{itemize}

The paper describes work which forms part of a "demand side maximisation" project and focuses on identifying typical usage profiles for households and then clustering them into a few archetypical profiles with similar kinds of customers grouped together. Differences between an individual household profile and that of others within the same group can be used to suggest energy usage behaviour changes to reduce overall electricity usage or to improve electrical efficiencies, possibly by time shifting particular appliance usage. In addition, particular groups (for example, large users during peak times) can be identified for targeting for reduction initiatives.

The work focuses on the daily variability of a household's electricity usage averaged over appropriate seasonal periods and similar types of days. Investigation of consumers' usage of electricity in order to determine similarities between types of consumers requires that the day's usage pattern is summarised in some way such that it can be compared with others. The "shape" of the usage pattern (e.g. little night usage, peak around breakfast, little usage during the day and then a peak during the evening period) needs to be determined. 

The purpose of the work is to test the applicability of applying the framework defined by \cite{figueiredo2005electric} to UK specific data and to identify possible enhancements or modifications to the framework in order to better fit the UK situation. In particular, the conclusion that a 2 stage process (Kohonen Self Organising Map and then Kmeans clustering) is the best approach to clustering the data is tested against the UK dataset.

\section{Background}

\begin{figure}[h]
\centering
\subfloat[Subfigure 1 list of figures text][Standard users]{
\includegraphics[width=0.2\textwidth]{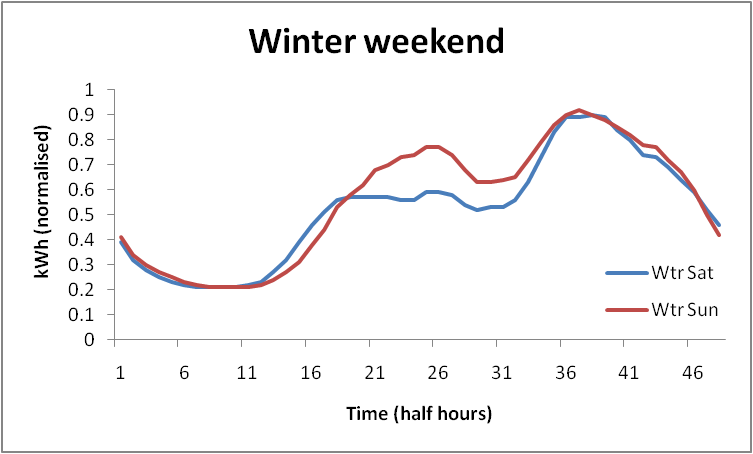}
\label{fig:subfig1}}
\subfloat[Subfigure 2 list of figures text][Economy 7 users]{
\includegraphics[width=0.2\textwidth]{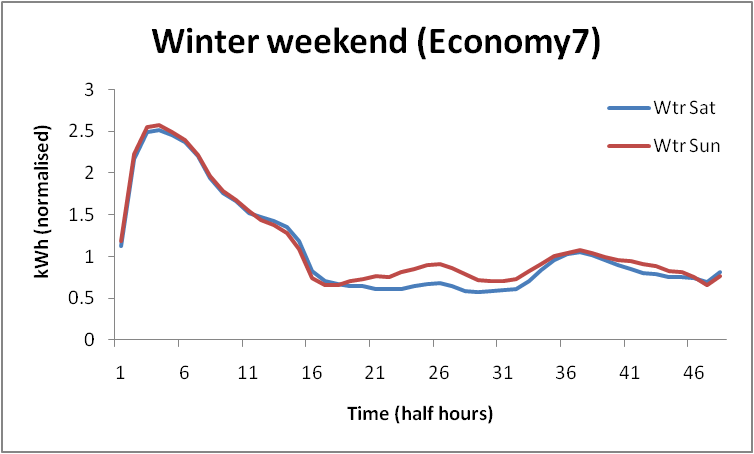}
\label{fig:subfig1}}
\caption{Example industry standard profiles}
\label{fig:industry}
\end{figure}

The Electricity industry defines a process \cite{association1997} for defining the details of eight standard usage profiles for the UK. The profiles take into account the season and the day of the week but only two refer to domestic properties.  As an example of the standard profiles, Figure \ref{fig:industry} shows the profiles for the winter for Saturday and Sundays, both for Economy 7 customers and non-Economy 7 customers, plotted as 48 half hourly readings over the day. Economy 7 is a tariff offer that provides much cheaper night time electricity (typically for 7 hours between 11pm and 8am) at the expense of slightly increased day time charges.

\cite{figueiredo2005electric} makes use of Portuguese data on 165 consumers, with readings taken at a 15 minute frequency, in order to validate the approach taken. The Figueiredo paper takes various differing clustering approaches and reaches the conclusion that a combination of Self Organised Maps (using a 10 x 7 grid), followed by a Kmeans algorithm to reach a final set of 9 clusters, is the best approach as measured by a "cluster quality" measure defined in the paper.

The Kmeans algorithm requires a number of clusters as an input parameter (n) and works by randomly selecting an initial n locations for the centres of the clusters. Each data point is then assigned to one of the centre locations by selecting the centre that is nearest to that data point. This definition of nearest differs in various algorithms but the Kmeans method uses Euclidean distance which is calculated for centre $c = (c_{1}, c_{2}, ..., c_{n})$ and point $p = (p_{1}, p_{2}, ..., p_{n})$ as
\begin{equation}
distance = \sqrt {\sum_{i} (c_{i} - p_{i})^2}
\end{equation}
Once all the data points are assigned to a centre, each collection of points is considered, the new centre of the allocated data points is calculated and the centre for that cluster is reassigned. The data points are then reallocated to their new nearest centre and the algorithm continues as before until no changes are made to the allocations of data points for an iteration. The method is highly dependent on the initial random allocation of centres \cite{jain1988algorithms}.

The Self Organising Map (SOM) is a neural network algorithm that can be used to map a high dimension set of data into a lower dimension representation. In the work presented in this paper, the mapping is to a 2 dimensional set of representations which are arranged in a hexagonal map. Each sample (mean load profile for a given household) is assigned to a position in the map depending on the closeness of the sample to the values on the nodes assigned to each position in the map (using a Euclidean measure of distance). Initially the nodes are assigned at random but, as samples are assigned to the nodes, the node incorporates the assigned data. Over time, the map produces an arrangement where similar samples are placed closely together and dissimilar are placed far apart \cite{kohonen2002self}.

The Figueiredo approach includes the following stages:
\begin{itemize}
\item	Cleaning of the data in order to cope with missing data and outliers in the data.
\item	Normalisation of the data to make differing readings comparable.
\item	Splitting of the data into typical types of day such as weekday, weekend, or holiday. The Figueiredo work concentrates on weekend and weekday split and seasonal split only.
\item	Creation of representative daily load profiles. Various approaches can be taken at this stage and Figueiredo uses the mean across all available days within the type of day and season.
\item Application of a number of clustering techniques in order to group the data into a pre-defined number of clusters and then the definition of a representative load profile for each cluster. A target number of clusters of nine is selected based on advice from, and general practice in, the Portuguese electricity industry together with some investigation on the quality of the clusters obtained when trying numbers of clusters between 6 and 14.
\item Calculation of the Mean Index Adequacy (MIA) as defined in \cite{chicco2003customer} in order to assess the comparative suitability of the generated clusters.
\end{itemize}

The data used in this study is from an area of Milton Keynes, UK which was developed in order to demonstrate various energy saving initiatives. The data was originally collected in 1988-91 by \cite{edwards1990low} but was then stored on floppy disks which deteriorated physically and some of the original data has been lost. The original data disks were rescued and, where possible, regenerated by Steve Pretlove of UCL and, more recently, by Alex Summerfield with the work detailed in \cite{summerfield2007milton}. The datasets have been made available in the UKERC data store.

\section{Methodology}
The approach detailed by Figueiredo has been applied to the UK data as closely as possible in order to assess the suitability of the framework application to the UK specific data. The individual steps in the process are detailed below.

\subsection{Cleaning}
Some of the UK data readings are missing readings for some hours of the day, either due to the way in which the data was recovered from floppy disks, or because of issues with the original collection of the data. For an initial view of the data, it was decided to omit all the days which contained a missing hourly reading. Alternative approaches to replacing some of the missing data making use of available data from a given similar day will be investigated in the future.

\subsection{Normalisation}
The UK data has been normalised within each day's readings by scaling all readings using the maximum hourly reading on the day set to 1. Thus all hourly readings are in the range 0-1. The effect of this normalisation is to focus on the shape of the usage pattern and not on the total usage. Two households with a similar shape (e.g. large early morning usage, little usage in the day and then medium evening usage) but with differing total usages (e.g. if one household is much larger than the other) will have the same normalised load profile once scaling is done against the largest hourly reading. The households will be clustered as similar in the further analysis whereas, depending on the way "similar" is defined, it might not be the intention to group these together (for example, if total electricity usage is to be the main differentiation between households).

\subsection{Stratifying the data}
The UK data was stratified using a split between weekend (Saturday and Sunday) and weekdays. It was further stratified into winter (the months of November, December, January, February, March, and April) and summer (the remaining months). With the variability of the UK climate, it may be more accurate to stratify the data based on daily temperatures rather than on the season and this will form the basis for future work. The data for winter weekends was arbitrarily chosen for further exploration as detailed in the remainder of this paper. Future work will concentrate on the other stratifications (e.g. summer weekday) which can be analysed in the same way and investigation will be done on how individual households are allocated to the same or different groupings as the season or type of day changes.

The Milton Keynes data has varying amounts of valid data for each household depending on the success of regeneration of the data after its rescue from floppy disks. The winter weekend data consists of between 25 and 111 valid days of readings for each of the households with a mean of 95 valid readings per household. Future analysis may suggest excluding some of the households with low values for valid data from the clustering but all have been included in this initial investigation.

\subsection{Creation of load profiles}
Each household has a representative average load profile generated by calculating the mean value for each hourly reading across all valid readings for the winter weekend. Other methods of calculating a representative profile could be adopted but this analysis has duplicated the approach taken with the Portuguese work.

\subsection{Application of clustering algorithms}
The Figueiredo approach compares the Kmeans algorithm with both a self-organised map (SOM) using a 3 x 3 grid and  also with a 2 stage process of first creating a SOM with 10 x 7 grid (i.e. 70 load diagrams) which are then clustered using the Kmeans algorithm. This approach has been replicated with the UK data although the volume of households is less (165 in Portugal, 93 in the UK) and hence the reduction in dimensions from the first SOM stage is not as great as with the Portuguese data.

The Kmeans clustering method relies on a random starting situation and requires the desired number of clusters to be input. In order to minimise the effects of the random starting point, the clustering algorithm was run 1000 times with differing random seeds. Examination of the results suggests that the large number of runs allows the same optimum solution to be found regardless of the starting random seed.

\begin{figure}
\includegraphics[width=0.45\textwidth]{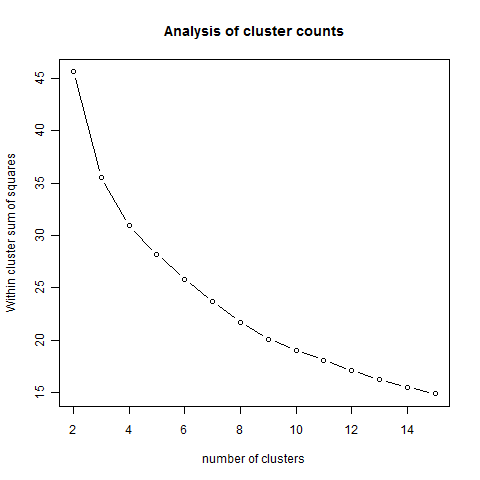}
\caption{Varying numbers of clusters input to Kmeans}
\label{fig:clusters}
\end{figure}

The within cluster sum of squares was calculated for each of the input numbers of clusters from 2 to 15. As the number of clusters increases, the total sum of squares will decrease (with the extreme example of each sample being in its own cluster with a total within cluster sum of squares being 0) and the graph (Figure \ref{fig:clusters}) can be examined to find an obvious "elbow" that denotes an appropriate number of clusters to select. The graph can be seen to be fairly uniform with no obvious elbows apart from that at 3 and possibly that at 9. In order to match the Portuguese work, the input value of 9 clusters was used for future analysis.

\subsection{Calculation of adequacy measure}
A measure is needed for assessing the quality of the clusters generated in order that the differing approaches can be compared. A good clustering scheme will create clusters where the members of a particular cluster are closely grouped but where the differences between members of differing clusters (or the representative profiles for the clusters) are well separated. A measure, Mean Index Adequacy (MIA), is defined in \cite{chicco2003customer} as
\begin{equation}
MIA = \sqrt{ \frac{1}{K} \sum\limits_{k=1}^K d^2 (r^{(k)}, C^{(k)})}
\label{equation-eqn1}
\end{equation}
where K clusters (k = 1..K) have been defined, $r^{(k)}$ is a load profile assigned to cluster k and $C^{(k)}$ is the calculated centre of the cluster k.

The distance between 2 load diagrams is defined as
\begin{equation}
d(l_i,l_j) = \sqrt{ \frac{1}{H} \sum\limits_{h=1}^H (l_i(h) - l_j(h))^2}
\end{equation}
where H is the number of individual readings in each load diagram (24 hourly readings) and $l_i(h)$ and $l_j(h)$ are the hth readings for two profiles, $l_i$ and $l_j$.

The MIA can be better described as 
\begin{equation}
MIA = \sqrt{ \frac{1}{K} \sum\limits_{k=1}^K \sum\limits_r d^2 (r^{(k)}, C^{(k)})}
\label{equation-eqn2}
\end{equation}
to signify the need to sum over all the distance calculations for each of the load profiles assigned to the given cluster (the distances between the load profiles and the cluster centre).

A lower value of MIA for a particular clustering solution signifies that the load profiles assigned to the calculated clusters are grouped closely together. A larger value denotes that the cluster members are widely dispersed and hence a low value for MIA is better and shows more compact clusters. The measure is useful as a comparison between differing clustering algorithms (where a lower value shows more compact clusters) but has little meaning as an absolute value.

The analysis work used R 2.12.2 running on a Samsung R580 laptop with Windows 7 Enterprise 64 bit operating system Service Pack 1. The laptop used an Intel i3 CPU (M350) running at 2.27 GHz and contained 3GB of memory.

\section{Results}
Differing clustering approaches were considered in order to explore the most appropriate for the UK data. 
\subsection{Kmeans}
\begin{figure}[h]
\centering
\subfloat[Subfigure 1 list of figures text][Cluster1]{
\includegraphics[width=0.16\textwidth]{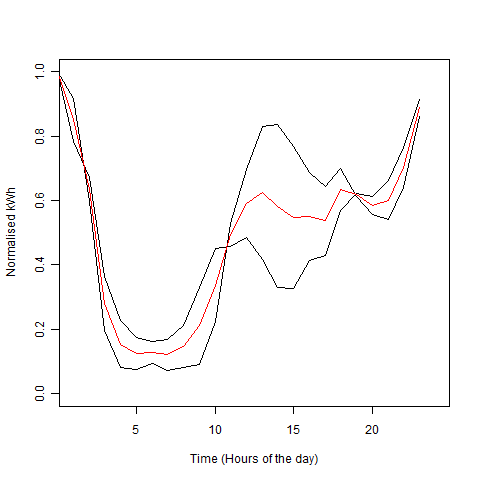}
\label{fig:subfig1}}
\subfloat[Subfigure 2 list of figures text][Cluster2]{
\includegraphics[width=0.16\textwidth]{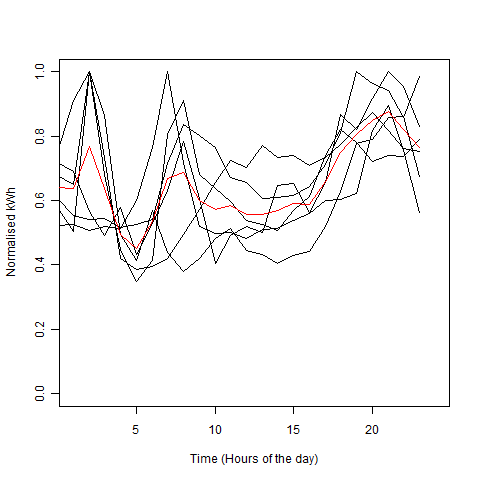}
\label{fig:subfig2}}
\subfloat[Subfigure 3 list of figures text][Cluster3]{
\includegraphics[width=0.16\textwidth]{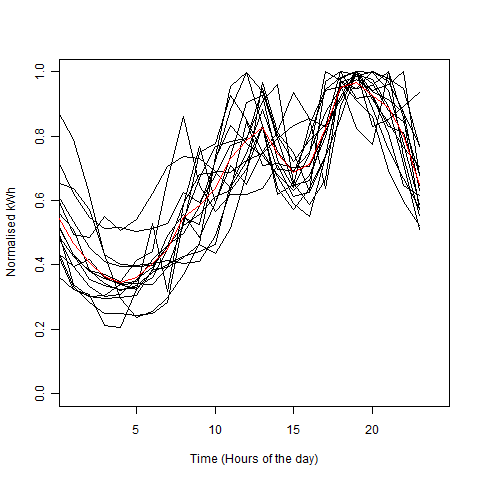}
\label{fig:subfig3}}
\qquad
\subfloat[Subfigure 4 list of figures text][Cluster4]{
\includegraphics[width=0.16\textwidth]{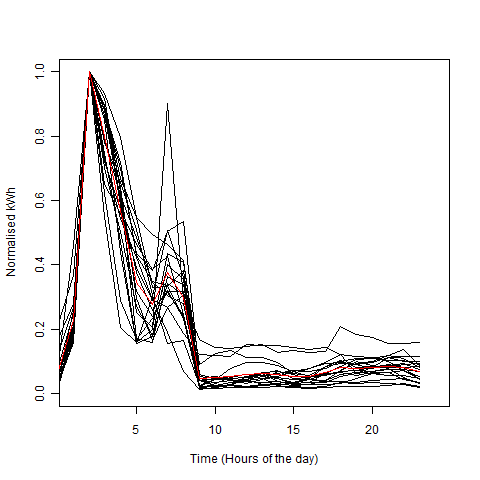}
\label{fig:subfig4}}
\subfloat[Subfigure 4 list of figures text][Cluster5]{
\includegraphics[width=0.16\textwidth]{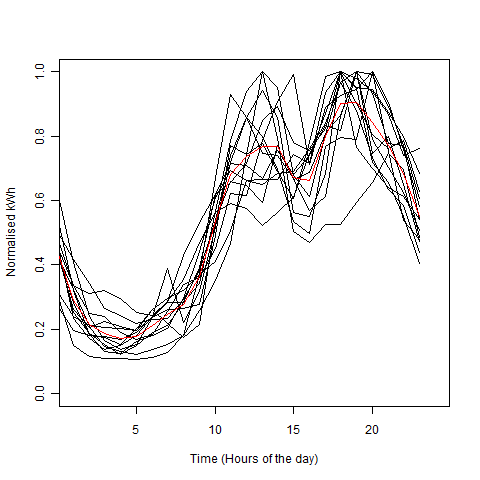}
\label{fig:subfig5}}
\subfloat[Subfigure 4 list of figures text][Cluster6]{
\includegraphics[width=0.16\textwidth]{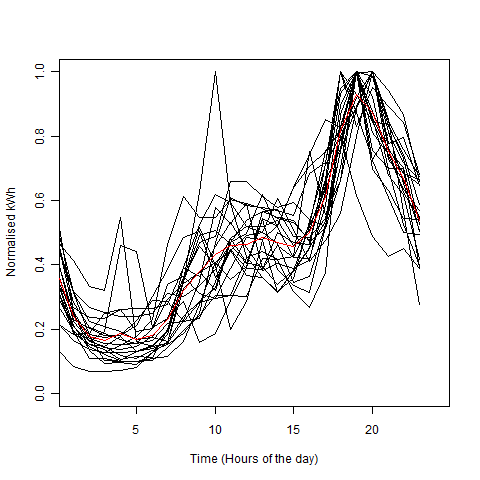}
\label{fig:subfig6}}
\qquad
\subfloat[Subfigure 4 list of figures text][Cluster7]{
\includegraphics[width=0.16\textwidth]{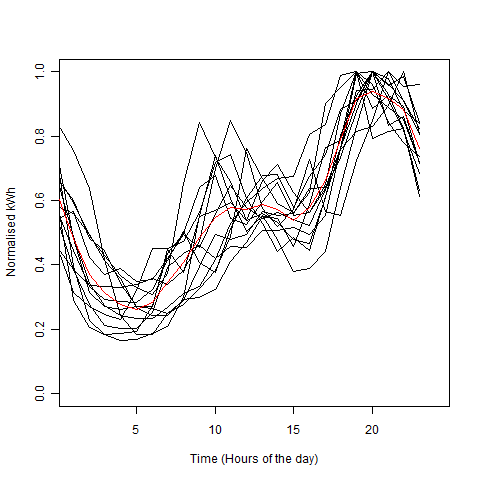}
\label{fig:subfig7}}
\subfloat[Subfigure 4 list of figures text][Cluster8]{
\includegraphics[width=0.16\textwidth]{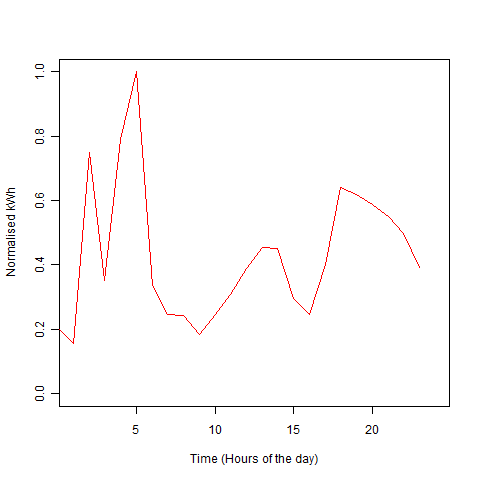}
\label{fig:subfig8}}
\subfloat[Subfigure 4 list of figures text][Cluster9]{
\includegraphics[width=0.16\textwidth]{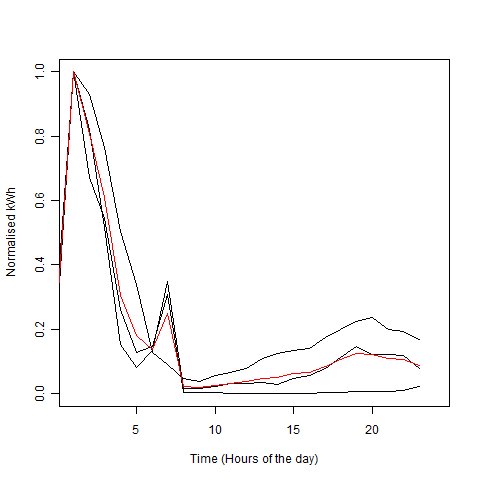}
\label{fig:subfig9}}
\caption{Clusters generated using Kmeans}
\label{fig:globfig}
\end{figure}

Initially the Kmeans clustering algorithm, with a target of nine clusters, was used to form the clusters. The clustering results using the Kmeans algorithm can be seen in Figure \ref{fig:globfig} where the black lines show the load profiles for the individual households allocated to the particular cluster and the red line shows the calculated representative profile for the cluster (the centroid). Where only one household is allocated to a cluster (e.g. as with "Cluster8"), the red line is overlaid on the black line.

The number of households allocated to each cluster by each technique are detailed in Table \ref{table-sizes}.

\begin{table}
\begin{center}
\renewcommand{\arraystretch}{1.3}
\caption{Size of clusters}
\label{table-sizes}
\begin{tabular}{|c|c|c|c|c|c|c|c|c|c|}
\hline
& 1 & 2 & 3 & 4 & 5 & 6 & 7 & 8 & 9\\
\hline
Kmeans: & 2 & 6 & 15 & 19 & 13 & 21 & 13 & 1 & 3 \\
\hline
SOM: & 2 & 6 & 15 & 22 & 12 & 9 & 12 & 1 & 14 \\
\hline
2 Stage: & 6 & 6 & 13 & 19 & 15 & 22 & 8 & 1 & 3 \\
\hline
\end{tabular}
\end{center}
\end{table}

\subsection{Self Organising Map}
\begin{figure}
\centering
\includegraphics[width=0.4\textwidth]{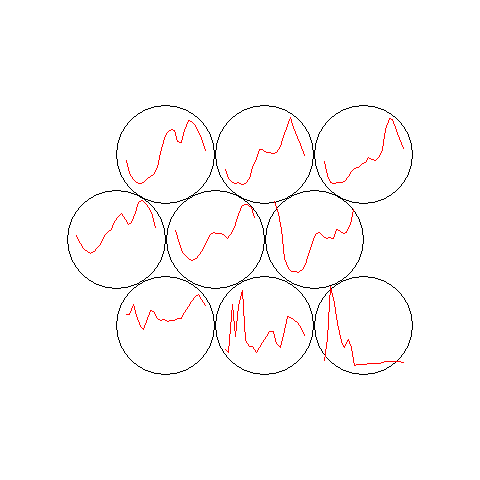}
\caption{Kohonen self organised map using 3 x 3 grid}
\label{fig:som}
\end{figure}

The Kohonen Self Organising Map algorithm was applied to the data using a hexagonal grid of 3 x 3 (i.e. 9 clusters). This creates the map of load profiles as shown in Figure \ref{fig:som}.

\begin{figure}[h]
\centering
\subfloat[Subfigure 1 list of figures text][Cluster1]{
\includegraphics[width=0.16\textwidth]{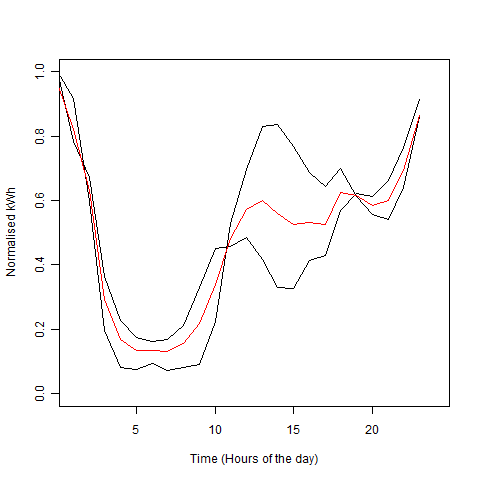}
\label{fig:subfig1}}
\subfloat[Subfigure 2 list of figures text][Cluster2]{
\includegraphics[width=0.16\textwidth]{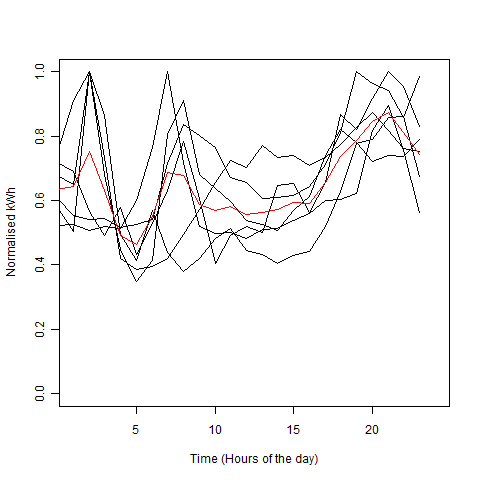}
\label{fig:subfig2}}
\subfloat[Subfigure 3 list of figures text][Cluster3]{
\includegraphics[width=0.16\textwidth]{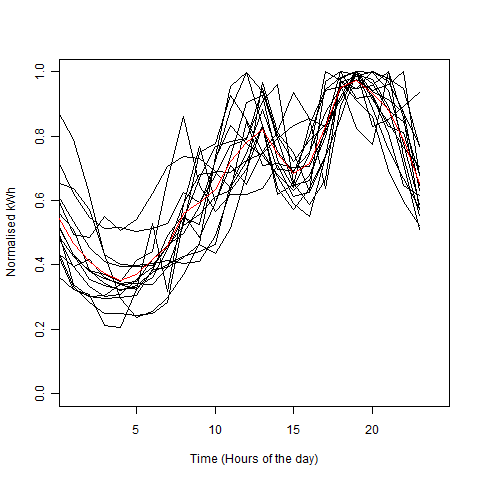}
\label{fig:subfig3}}
\qquad
\subfloat[Subfigure 4 list of figures text][Cluster4]{
\includegraphics[width=0.16\textwidth]{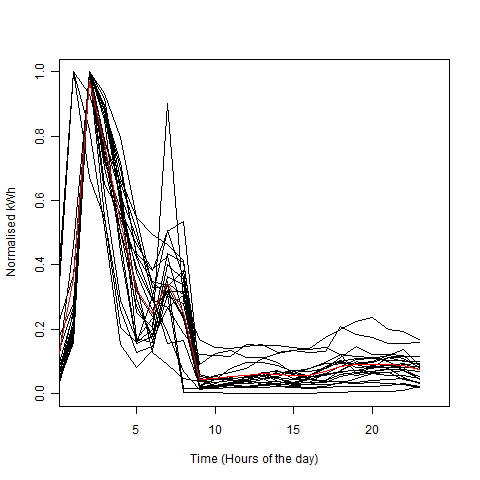}
\label{fig:subfig4}}
\subfloat[Subfigure 4 list of figures text][Cluster5]{
\includegraphics[width=0.16\textwidth]{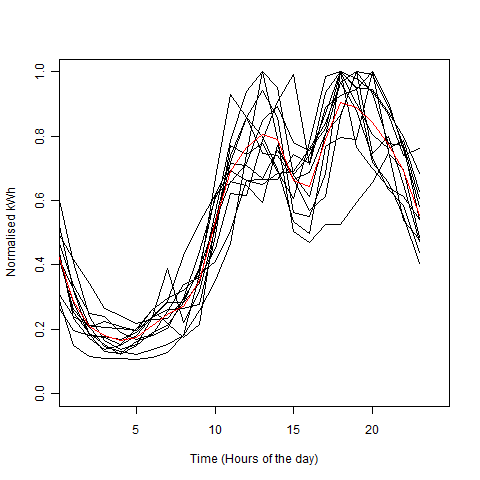}
\label{fig:subfig5}}
\subfloat[Subfigure 4 list of figures text][Cluster6]{
\includegraphics[width=0.16\textwidth]{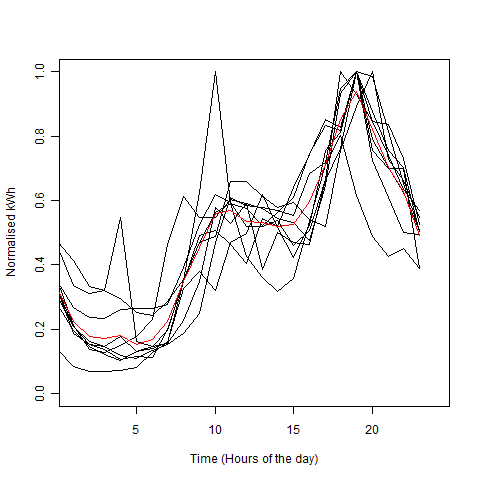}
\label{fig:subfig6}}
\qquad
\subfloat[Subfigure 4 list of figures text][Cluster7]{
\includegraphics[width=0.16\textwidth]{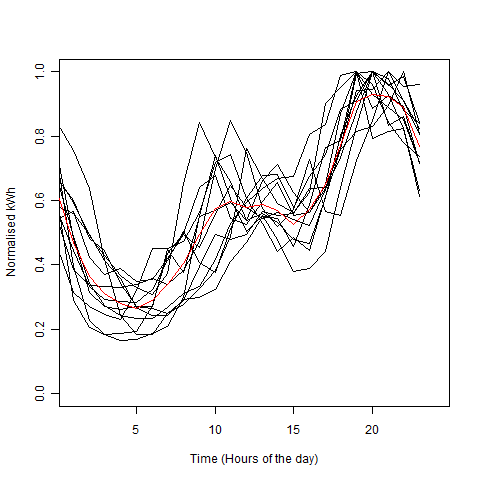}
\label{fig:subfig7}}
\subfloat[Subfigure 4 list of figures text][Cluster8]{
\includegraphics[width=0.16\textwidth]{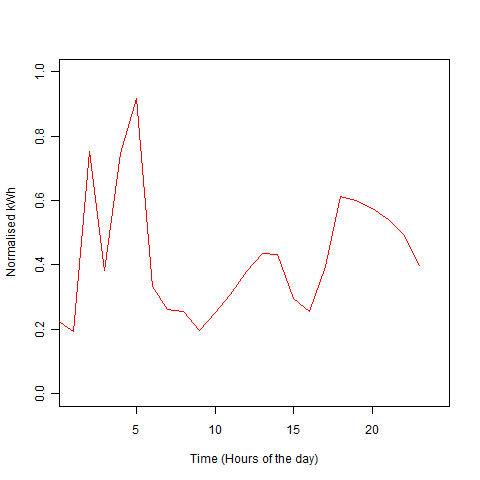}
\label{fig:subfig8}}
\subfloat[Subfigure 4 list of figures text][Cluster9]{
\includegraphics[width=0.16\textwidth]{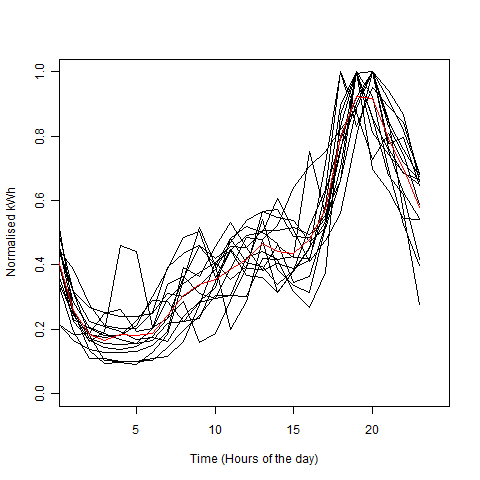}
\label{fig:subfig9}}
\caption{Clusters generated using Kohonen Self Organising Maps}
\label{fig:globfig2}
\end{figure}

Plotting the household load profiles alongside the calculated cluster centres produces the results in Figure \ref{fig:globfig2} with the numbers of households allocated to each cluster listed in Table \ref{table-sizes}. The clusters are numbered randomly and the order in the displays has been modified in order to match the Kmeans clusters as far as possible. The match between the generated clusters is fairly obvious with the exception of "Cluster9" which is significantly different.

\subsection{Two stage process}

The hypothesis in \cite{figueiredo2005electric} is that the application of a Kohonen Self Organising Map algorithm to the data in order to create 70 (10 x 7) clusters in a hexagonal grid followed by the application of the Kmeans algorithm to the SOM output produces the best clusters as defined by the MIA measure. This work was replicated using the UK data although the number of households is lower than that used in the Portuguese work. 

\begin{figure}
\centering
\includegraphics[width=0.4\textwidth]{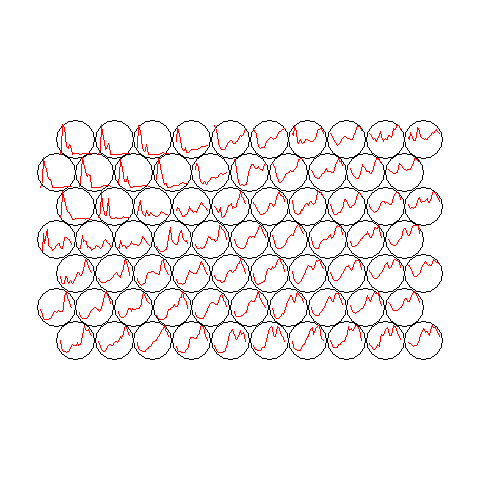}
\caption{Kohonen self organised map using 10 x 7 grid}
\label{fig:intermediate}
\end{figure}

\begin{figure}[h]
\centering
\subfloat[Subfigure 1 list of figures text][Cluster1]{
\includegraphics[width=0.16\textwidth]{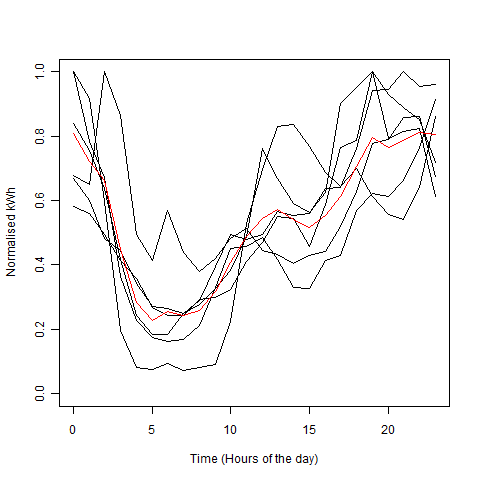}
\label{fig:subfig1}}
\subfloat[Subfigure 2 list of figures text][Cluster2]{
\includegraphics[width=0.16\textwidth]{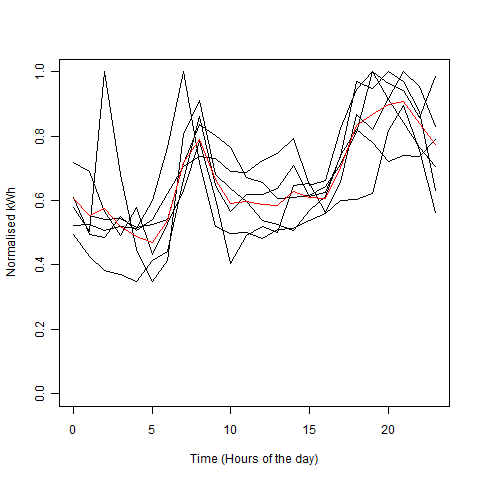}
\label{fig:subfig2}}
\subfloat[Subfigure 3 list of figures text][Cluster3]{
\includegraphics[width=0.16\textwidth]{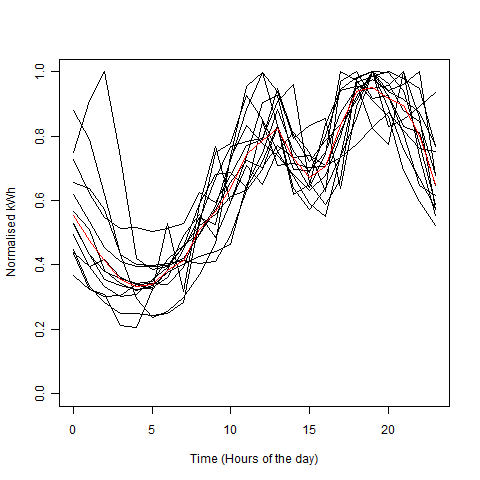}
\label{fig:subfig3}}
\qquad
\subfloat[Subfigure 4 list of figures text][Cluster4]{
\includegraphics[width=0.16\textwidth]{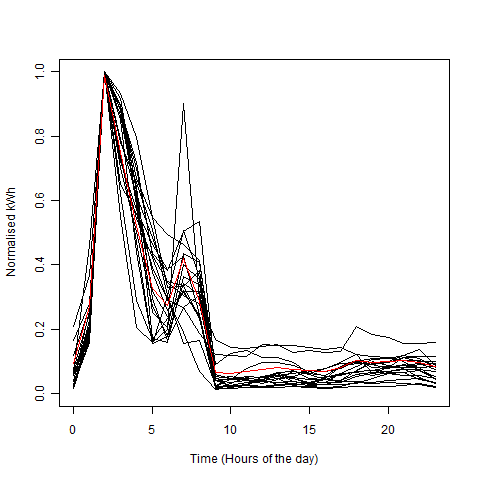}
\label{fig:subfig4}}
\subfloat[Subfigure 4 list of figures text][Cluster5]{
\includegraphics[width=0.16\textwidth]{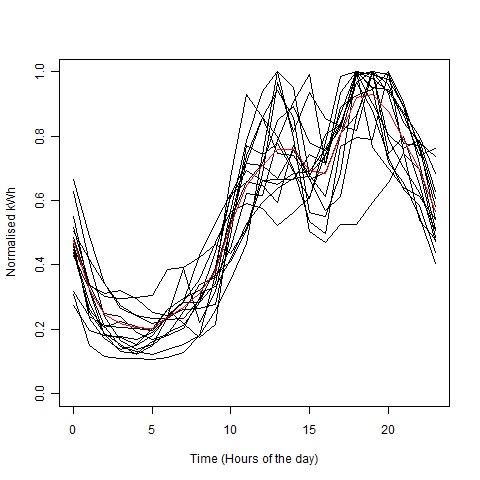}
\label{fig:subfig5}}
\subfloat[Subfigure 4 list of figures text][Cluster6]{
\includegraphics[width=0.16\textwidth]{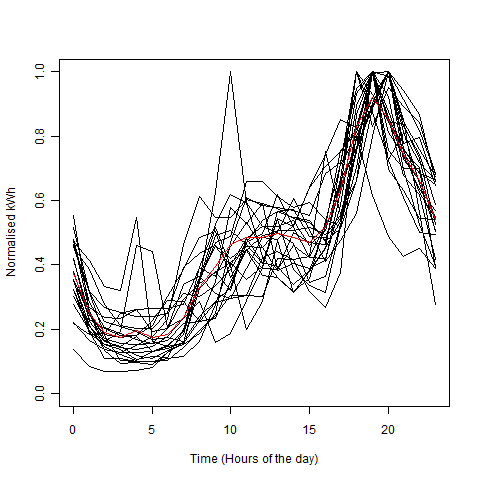}
\label{fig:subfig6}}
\qquad
\subfloat[Subfigure 4 list of figures text][Cluster7]{
\includegraphics[width=0.16\textwidth]{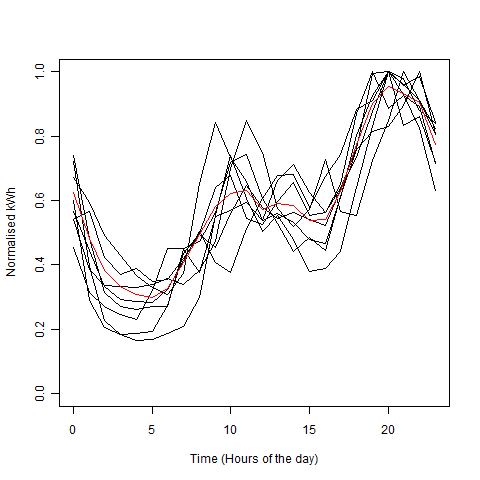}
\label{fig:subfig7}}
\subfloat[Subfigure 4 list of figures text][Cluster8]{
\includegraphics[width=0.16\textwidth]{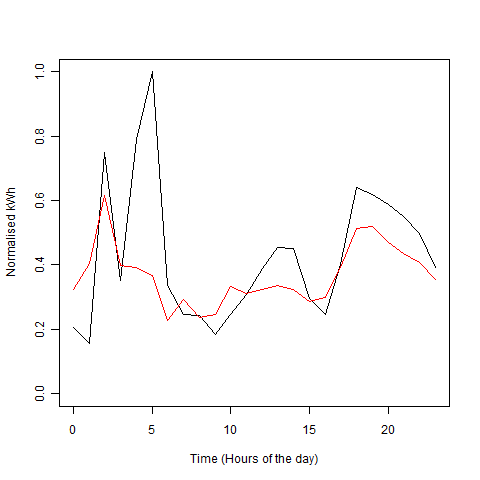}
\label{fig:subfig8}}
\subfloat[Subfigure 4 list of figures text][Cluster9]{
\includegraphics[width=0.16\textwidth]{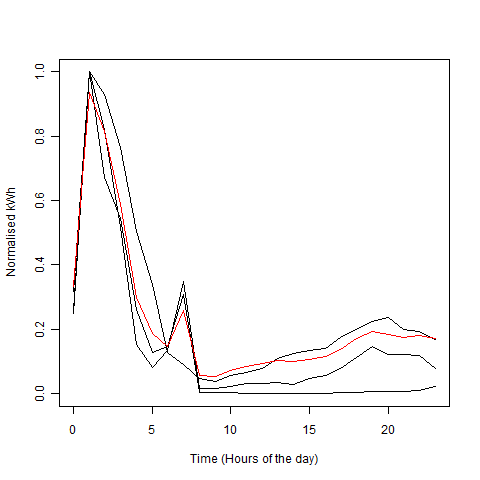}
\label{fig:subfig9}}
\caption{Clusters generated using the 2 stage process}
\label{fig:globfig3}
\end{figure}

The intermediate map generated by the SOM is shown at Figure \ref{fig:intermediate}. The intermediate load profiles shown are then input to the Kmeans algorithm in order to generate nine final clusters. The original allocation of household load profiles to the intermediate SOM and thence to the final clusters is then examined in order to determine the number of households in each final cluster and to allow for plotting of the final cluster profiles alongside the households allocated to that cluster. Again the order of the generated clusters has been altered to match the Kmeans generated clusters as closely as possible. The results are shown in Figure \ref{fig:globfig3} with the number of households allocated to each cluster detailed in Table \ref{table-sizes}.

\subsection{Comparison of clustering techniques}

The MIA figures for each clustering approach were calculated and are listed in Table \ref{table-tab1} with a lower figure denoting more compact clusters. The results show the best algorithm for clustering (as measured by MIA) is the Kmeans approach. 

The MIA measure is very sensitive to the few profiles which differ from the profile for the generated cluster to which they are allocated. This sensitivity may detract from the MIA as a good measure of clustering success as, whilst most of the households may be well clustered, a single household profile allocated to one cluster rather than another can greatly increase the MIA value and hence reduce the measured effectiveness of the clustering. It is proposed in future work to examine alternative clustering measures and to assess the sensitivity of the measures to a few profiles which are difficult to allocate to clusters. 

\begin{table}
\begin{center}
\renewcommand{\arraystretch}{1.3}
\caption{MIA calculations}
\label{table-tab1}
\begin{tabular}{|c|c|c|c|}
\hline
& Kmeans & Kohonen SOM & 2 stage process\\
\hline
MIA value: & 0.3050533 & 0.3166297   & 0.3205487\\
\hline
\end{tabular}
\end{center}
\end{table}

The graphs showing the generated clusters and the households that are allocated to each cluster show that each technique produces some clusters which appear visually to be very similar but also some clusters that vary widely. In particular the "Cluster9" is significantly different for the various clustering techniques. The numbers of households allocated to each cluster can be seen to vary and this demonstrates that the clustering techniques will have differing levels of success in generating the best split into clusters.

\section{Conclusions and future work}
The work demonstrates that UK domestic load profiles can be successfully clustered and the visual impression from the cluster representative profiles is of very differing shapes of usage. In particular, the load shapes differ significantly from the standard domestic profiles used by the industry which are only differentiated by Economy 7 usage (see Figure \ref{fig:industry}). This shows that the application of appropriate clustering techniques will allow for more accurate differentiation between household usage patterns than that currently published by the industry and will lead to more accurate representative profiles which can be used for demand aggregation, supply side planning, marketing and other purposes.

The selection of nine as the target number of clusters reflects the decision taken in Portugal based on input from the Portuguese electricity industry and from analysis of the cluster measures for differing numbers of target clusters. The evidence for selecting nine clusters for the UK winter weekend data is weak and more investigation of an appropriate target number of clusters appropriate to the UK data is planned.

The work undertaken in Portugal using Portuguese data concluded that using a two-stage process of building a Self Organising Map and then applying a Kmeans clustering algorithm was the most effective in generating well distinguished clusters as measured by the MIA measure. The UK data does not support this conclusion and the best MIA figure is from the simple application of the Kmeans algorithm. In fact, it was found that the SOM technique alone provided better results (as measured by the MIA measure) than the two-stage process.

Analysis has been concentrated on the winter weekend data and other slices across the data may show differing results. In particular it may be found that households are clustered together differently for different types of day (by season or weekend/weekday) and year long stable clusters, with the same members for each season, may not be identifiable. Future work is planned to investigate this further.

The MIA measure of the quality of the generated clusters is very sensitive to a few households which are hard to allocate and differing measures of cluster quality will be investigated in the future. 

The normalisation used in the exercise has the effect of comparing shapes of usage but not absolute values of usage and a clustering approach that differentiates a household using much more electricity from another using less may be required (depending on the use to be made of the clusters found). The appropriateness of the normalisation is related to the definition of "similar" users which will be explored in future work.

\section*{Acknowledgements}
This data was accessed through the UK Energy Research Centre Energy Data Centre (UKERC-EDC). Our acknowledgements to the Building Research Establishment, which provided access to the original 1990 data set from Milton Keynes Energy Park, and to Bartlett School of Graduate Studies, University College London for processing and cleaning the raw data.

This work is possible thanks to EPSRC grant reference EP/I000496/1.

\bibliographystyle{plain}
\bibliography{References}

\end{document}